\documentclass[onecolumn]{aastex62}
\pdfoutput=1 
\usepackage{amsmath,amstext}
\usepackage[T1]{fontenc}
\usepackage{apjfonts} 
\usepackage[figure,figure*]{hypcap}
\usepackage{subfigure}
\usepackage{epsfig}
\usepackage{longtable}
\usepackage{isotope}

\newcommand{\rev}[1]{#1}
\newcommand{\cutt}[1]{\textcolor{blue}{}}

\newcommand{\Ms}{{\ensuremath{{M}_{\odot} }}}

\newcommand{\GIZMO}{\texttt{GIZMO}}
\newcommand{\NumDen}{cm$^{-3}$}
\newcommand{\TNG}{\texttt{IllustrisTNG}}
\newcommand{\Mach}{$\mathcal{M}$}

\shorttitle{Primordial Turbulence}
\shortauthors{Chen, Ho, and Tung.}

\begin{document}

\title{Formation of Supersonic Turbulence in the Primordial Star-forming Cloud}

\author[0000-0002-4848-5508]{Ke-Jung Chen}
\affiliation{Institute of Astronomy and Astrophysics, Academia Sinica, No.1, Sec. 4, Roosevelt Rd., Taipei 10617, Taiwan, R.O.C.}
\affiliation{Heidelberg Institute for Theoretical Studies, Schloss-Wolfsbrunnenweg 35, Heidelberg 69118,
Germany}

\correspondingauthor{Ke-Jung Chen}
\email{kjchen@asiaa.sinica.edu.tw}

\author[0000-0002-2316-1371]{Meng-Yuan Ho}
\affiliation{Institute of Astronomy and Astrophysics, Academia Sinica, No.1, Sec. 4, Roosevelt Rd., Taipei 10617, Taiwan, R.O.C.} 
\affiliation{Department of Physics, National Taiwan University, No.1, Sec. 4, Roosevelt Rd.,  Taipei 10617, Taiwan, R.O.C.} 

\author[0009-0000-9401-5470]
{Pei-Cheng Tung}
\affiliation{Institute of Astronomy and Astrophysics, Academia Sinica, No.1, Sec. 4, Roosevelt Rd., Taipei 10617, Taiwan, R.O.C.} 
\affiliation{Department of Physics, National Taiwan University, No.1, Sec. 4, Roosevelt Rd.,  Taipei 10617, Taiwan, R.O.C.}

\begin{abstract}
We present new simulations of the formation and evolution of the first star-forming cloud within a massive minihalo of mass of $1.05 \times 10^7\,\Ms$, carried out using the \GIZMO\ code with detailed modeling of primordial gas cooling and chemistry. Unlike previous studies that simulated the formation of the first stars within a smaller cosmological boxsize of \rev{$\sim 0.3-2$} Mpc, our work adopts initial conditions from the large-scale cosmological simulations, \TNG\ spanning $\sim 50$ Mpc to study the formation of primordial clouds that give birth to the first stars. We increase the original resolution of \TNG\ by a factor of $\sim10^5$ using a particle-splitting technique, achieving an extremely high resolution that allows us to resolve turbulence driven by gravitational collapse during early structure formation. We find that strong supersonic turbulence with a characteristic Mach number of $\sim 5.2$ naturally develops within the collapsing halo. This turbulence efficiently stirs the gas, promoting fragmentation of the star-forming cloud into multiple dense clumps. Among them, we identify a gravitationally bound core with a mass of $8.07\,\Ms$ and a size of $0.03$ pc, which exceeds its local Jeans mass and is on the verge of collapsing into a star. Our results indicate that supersonic turbulence may be common in primordial halos and can play a crucial role in cloud-scale fragmentation, \rev{ providing an alternative channel to form less massive first stars and strengthens the argument of lowering characteristic mass for the first stars found in previous studies.}

\end{abstract}


\section*{Introduction}
Turbulence is ubiquitous in the astrophysical environment, spanning from stellar to galactic and even larger scales \citep{larson79turb,shore92,turb1,kowal14}. In the interstellar medium (ISM), turbulence is primarily driven by feedback from massive stars, such as stellar winds and supernovae (SNe) \citep{kru16feedback,kim17sn,koo20sn,bac20sn}. On galactic and cluster scales, energy injection from active galactic nuclei (AGNs) and large-scale outflows further stir the gas, contributing to turbulence in galaxy clusters \citep{par10cluster,ro11cluster,hu22turb}. This turbulent motion plays a fundamental role in regulating the baryonic cycle of cosmic gas \citep{ferri01galaxy,low04rmp,fed21_turb}.

A natural question arises: Could turbulence exist even before the first stars so-called Population~III (Pop~III) stars formed? According to the standard cosmological model, the Universe emerged from the Big Bang followed by a brief period of inflation, imprinting quantum fluctuations onto spacetime. These fluctuations seeded the initial matter density perturbations, which later grew via gravitational instability into the cosmic web of filaments and halos. The so-called dark matter minihalos, with masses of $\sim 10^5$–$10^6\,\Ms$, are thought to be the cradles of Pop~III stars \citep{Tegmark_1997,abn02,bl04c,y08,hir15,ks23}. At that epoch, stellar feedback was absent; the evolution was governed solely by gravitational collapse and the hydrodynamics of primordial gas accreting into dark matter potential wells.

As structure formation proceeded, the collapse of gas into these early minihalos could induce turbulent motions—so-called gravitational turbulence—analogous to what is seen in present-day star-forming molecular clouds \citep{klessen00turb,henneb21_gturb,fensch23_gturb}. However, previous simulations of Pop~III star formation lacked sufficient resolution to capture the full turbulence cascade across the entire minihalo \citep{turk09,stacy10,hir13,greif14,chen15}. Instead, these studies typically employed hierarchical zoom-in techniques to resolve the central star-forming regions down to sub-parsec or even AU scales, but the larger-scale environment of the halo remained under-resolved, making it difficult to capture the emergence of turbulence during halo assembly. If supersonic turbulence arises during this phase, \rev{it can fragment the entire star-forming cloud into multiple clumps, which subsequently collapse to form stars, thereby influencing the initial mass function (IMF) of Pop~III stars}
\citep{tang24}. To investigate this possibility, we perform high-resolution cosmological zoom-in simulations using the \GIZMO\ code, starting from initial conditions extracted from the publicly available \TNG\ project \citep{2018Springel}. We increase the original mass and spacial resolution in \TNG\ by a factor of $\sim10^5$ via a particle-splitting technique, allowing us to follow the formation of turbulent gas during the assembly of minihalo with unprecedented detail.

The structure of this paper is as follows. In Section 2, we describe the methodology of our numerical simulations. In Section 3, we present our results and demonstrate the formation of supersonic turbulence within a minihalo. Section 4 discusses the impact of strong turbulence on the formation of Pop~III stars. Finally, we summarize our findings in Section 5.

\section{Methods}
\subsection{\GIZMO\ Code}

We perform our simulations using the mesh-free hydrodynamics code \GIZMO\ \citep{Hopkins_2015, Hopkins_2018_cooling}. \GIZMO\ is an open-source code built upon the widely used cosmological code \texttt{GADGET-2} \citep{Springel2003}, utilizing its domain decomposition and N-body algorithms. Our simulations adopt the Meshless Finite-Mass (MFM) method, in which each particle serves as a mesh-generating point that defines a fluid volume. The physical properties associated with each particle represent the local state of the fluid, and the hydrodynamic equations are solved by integrating over the particle-defined volumes.

To model primordial gas cooling and chemistry, we couple \GIZMO\ with the \texttt{GRACKLE} package \citep{2017MNRAS.466.2217S_Grackle}. \texttt{GRACKLE} provides a detailed treatment of the non-equilibrium chemistry and cooling processes relevant for primordial gas, including a network of twelve species: $\mathrm{H,\, H^+,\, H^-,\, D,\, D^+,\, HD,\, H_2,\, H_2^+,\, He,\, He^+,\, He^{++},}$ and $\mathrm{e^-}$. The cooling processes account for collisional excitation and ionization, radiative recombination, free-free emission, and molecular cooling via $\mathrm{H_2}$ and HD, including line emission, formation heating, and collision-induced emission. The chemistry and cooling networks are fully coupled with the hydrodynamics, ensuring a self-consistent treatment of the primordial gas physics throughout the simulations.

\subsection{Initial conditions from \TNG}

Our \GIZMO\ simulations adopt initial conditions from the state-of-the-art cosmological simulations of the \TNG\ project, which evolves the universe from $z = 127$ to $z = 0$ with large-volume gravo-magnetohydrodynamical calculations \citep{2018Pillepich, 2018Nelson, 2018Naiman, 2018Springel, 2018Marinacci, 2019Nelson, 2019Pillepich}. Specifically, we use the highest resolution data from the TNG50 simulation, which covers a 50~cMpc box. Minihalos are selected from the TNG50 halo catalog using the {\footnotesize SUBFIND} algorithm \citep{2001Springel}, which identifies gravitationally bound structures. To preserve the surrounding cosmic environment, the entire \textsc{FoF} (Friends-of-Friends) group associated with the selected minihalo is also included. 
We identify a minihalo progenitor from the TNG50 catalog at \( z = 20 \) and extract a cubic volume of \( \sim 14.7^3 \)~ckpc\(^3\) encompassing this halo. This volume is then mapped onto \GIZMO\ as the initial condition for our simulations.

The original mass resolution in TNG50 is $\sim 4.56\times10^{5}\,\Ms$ for dark matter particles and $\sim 8.38\times10^{4}\,\Ms$ for gas particles. At this resolution, minihalos are represented by only a few tens of particles, insufficient to resolve the turbulence formation within the halo. To achieve the necessary mass resolution, we apply a super-Lagrangian particle refinement method \citep{Hopkins_2015}, in which particles are progressively split to increase resolution, similar to methods employed in \cite{alc21_split,2024Tung, 2024Ramesh_1, 2024Ramesh_2}. The refinement is implemented during the first five million years of the simulation, a phase when self-gravity of both gas and dark matter, coupled with hydrodynamics, is actively shaping structure formation. Particle splitting is applied gradually, refining the mass resolution as collapse proceeds. After approximately 5~Myr, the target mass resolution of $\sim 0.19\,\Ms$ for gas and $\sim 80.88\,\Ms$ for dark matter is reached, and dense gas structures begin to emerge. At this point, we activate the primordial chemistry and cooling networks provided by \texttt{GRACKLE} and continue the evolution of the system. 

\rev{We refine all particles/meshes uniformly instead of using adaptive mesh refinement (AMR) method because AMR uses a dynamically refined mesh to increase resolution in specific regions of interest, such as shock fronts or high-density star-forming regions. This approach requires predefined refinement criteria to determine where the mesh should be refined. However, in highly turbulent environments—such as stellar convection zones or giant molecular clouds—turbulence is pervasive throughout the domain. Capturing such widespread turbulence would require refinement across most of the simulation volume, reducing the computational efficiency.
Moreover, turbulence evolves from large driving scales to small dissipation scales, spanning a wide range of gas densities and velocities. This multiscale nature makes it inherently challenging to define refinement criteria that can reliably capture the formation and evolution of turbulent structures. As a result, AMR is less effective for simulations that aim to model fully developed turbulence across the entire computational domain.
}


The simulation is evolved from $z=20.05$ to $z=18.78$, corresponding to approximately 17.5Myr of physical time. By the end of the run, a massive minihalo with a total mass of $1.05\times10^7 \,\Ms$ has formed, containing $1.34\times10^6 \,\Ms$ of gas and $9.16\times10^6 \,\Ms$ of dark matter, with a virial radius of 336.48 pc. Since this study is to study turbulence formation during minihalo assembly, we do not implement sink particles for star formation. At late times, the collapse of dense gas at small scales leads to a dramatic reduction in the simulation timestep because of the Courant–Friedrichs–Lewy (CFL) condition \citep{nick18_cfl}, ultimately causing the simulation to terminate.

\section{Results}
\subsection{Gas accretion during the formation of a minihalo}

\begin{figure}[tbh]
\centering
\includegraphics[width=\columnwidth]{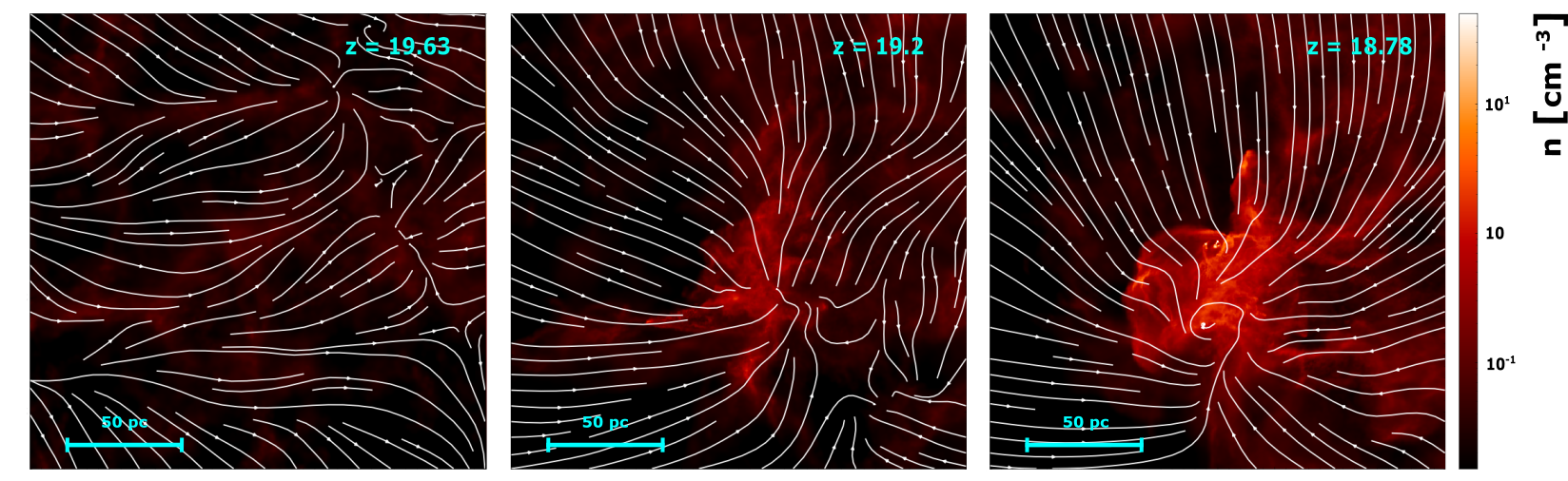}
\caption{Density evolution during the assembly of a minihalo. Panels from left to right show snapshots at different stages of the evolution. The streamlines illustrate the direction of gas flow. The initially diffuse and featureless gas distribution rapidly evolves into a concentrated gas cloud as the minihalo forms. Filamentary and clumpy structures emerge within the cloud, likely driven by anisotropic accretion flows of primordial gas.}
\label{fig:evolve}
\end{figure}

We first present the evolution of the gas distribution during the assembly of a minihalo in Figure~\ref{fig:evolve}. At $z=19.63$, the minihalo is in the early stages of assembly, and the gas remains distributed across separate regions. However, the velocity streamlines reveal a tendency for gas accumulation at the convergence points, indicating the presence of a strong gravitational attractor associated with small dark matter structures. By $z=19.2$, the clustering gas concentrates toward the halo center, showing a rapid accretion of gas along filaments into the halo. At $z=18.78$, a dense gas cloud has formed, exhibiting fine, thin structures within it. The velocity streamlines converge toward the center, and circular patterns indicate solenoidal (rotational) motion in the accreting flows. This evolution demonstrates that the gas accretion is highly anisotropic and inhomogeneous, resulting in clumpy structures, which are likely shaped by tidal forces from the assembling dark matter halo.

\begin{figure}[tbh]
\centering
\includegraphics[width=\columnwidth]{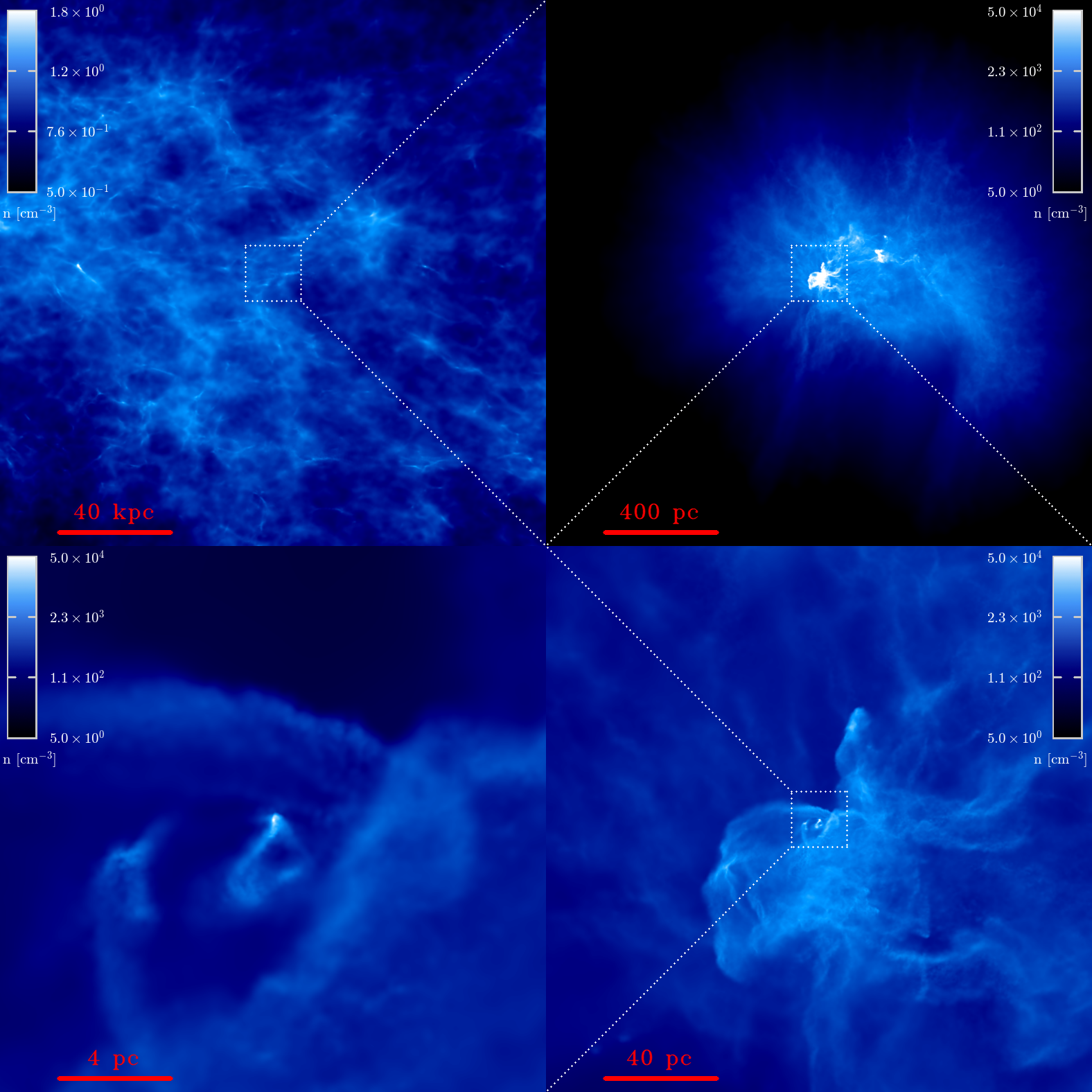}
\caption{Morphology of a primordial minihalo at $z=18.78$. The series of panels shows successive zoom-ins of the gas density from a scale of 40 kpc down to the inner 4 pc of the targeted halo. Clumpy structures become increasingly prominent at smaller scales. In the 4 pc panel, the central region exhibits an elongated dense clump surrounded by a tail of circularly streaming gas, highlighting the complex, anisotropic dynamics within the collapsing core. }
\label{fig:zoom}
\end{figure}

Figure~\ref{fig:zoom} shows zoom-in views of the minihalo at the end of the simulation. The distribution of gases within the virial radius is highly inhomogeneous, with thin and fragmented structures extending from the scales of $\sim 100$~ pc to $\sim 1$~ pc. These large-scale inhomogeneities cascade down to cloud scales, where star formation eventually occurs. This suggests that primordial star-forming clouds are likely to be intrinsically clumpy.

\subsection{Physical Properties of Primordial Turbulence}

\begin{figure}[tbh]
\centering
\includegraphics[width=\columnwidth]{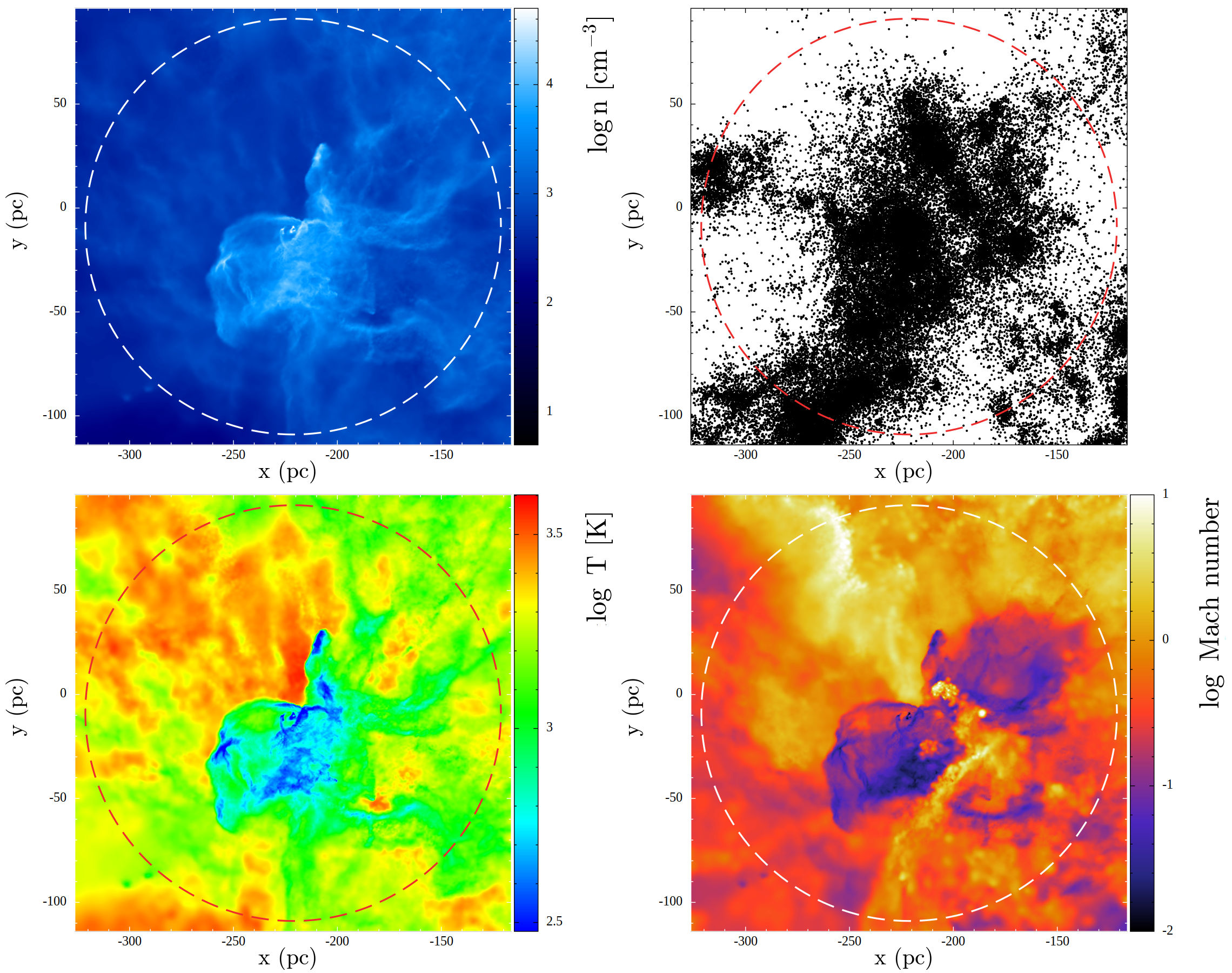}
\caption{Physical properties of a primordial halo. Panels show the gas density, dark matter distribution, gas temperature, and \Mach\ at the end of the simulation. The dashed circle marks the inner region of the halo with a radius of 100 pc. The gas density morphology closely traces the underlying dark matter structure, with dense gas regions coinciding with concentrations of dark matter. The central high-density gas exhibits lower temperatures, primarily due to efficient cooling by molecular hydrogen.}
\label{fig:4in1}
\end{figure}

We show the distributions of gas density, dark matter, gas temperature, and gas Mach number (\Mach) within the minihalo at the end of the simulation in Figure~\ref{fig:4in1}. The gas distribution generally traces that of dark matter, with most of the gas residing near the halo center. The high-density regions of gas correspond well to lower-temperature areas, as a result of molecular hydrogen cooling.
From the \Mach\ distribution, the halo is filled with a mixture of subsonic and supersonic gas. Supersonic accretion flows dominate around $r \sim 5-100$ pc. The presence of supersonic turbulence stirs the gas in the halo center, creating clumpy structures that can significantly influence subsequent star formation inside these dense clumps.

\begin{figure}[tbh]
\centering
\includegraphics[width=0.6\columnwidth]{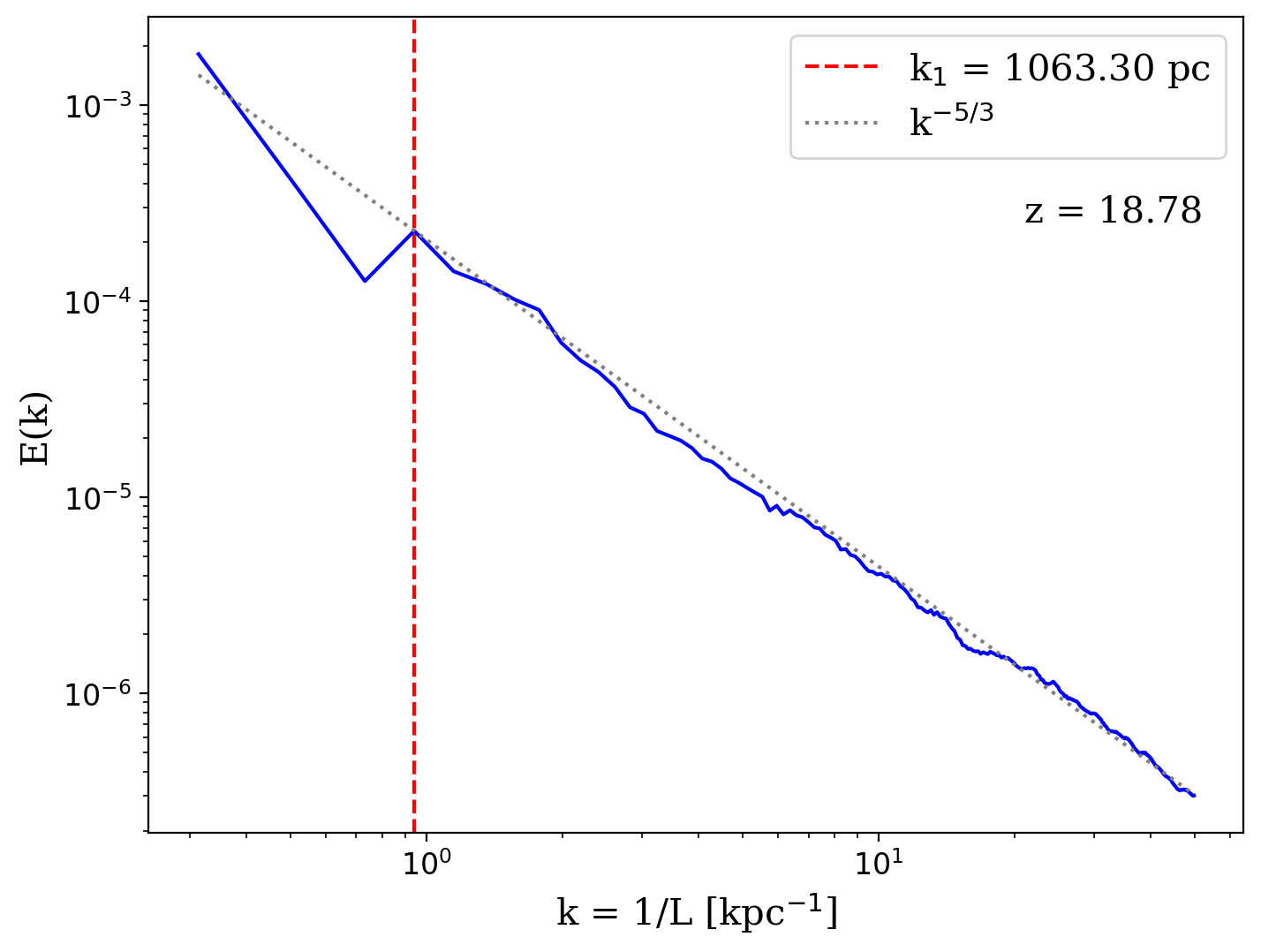}
\caption{Kinetic energy power spectrum of gas in the halo. The energy spectrum has a turning point at $k \approx k_1$, corresponding to a physical scale of 1063 pc—approximately three times the halo’s virial radius. When $k > k_1$, the spectrum follows the Kolmogorov power-law scaling of $k^{-5/3}$, indicative of a turbulent inertial cascade, down to $k \approx 50$, which corresponds to a physical scale of about 2 pc.}
\label{fig:spec}
\end{figure}

In addition, we quantify the gas kinematics of the halo by showing the power spectrum of the gas kinetic energy in Figure~\ref{fig:spec}. The spectrum follows a power-law profile consistent with the Kolmogorov spectrum \citep{1991_Kolmogorov}. The Kolmogorov theory divides turbulence into three characteristic scales: the energy-driving scale, the inertial scale, and the dissipation scale.
As shown in Figure~\ref{fig:spec}, we identify a turning point at small wavenumbers $k\sim 1\,\mathrm{kpc}^{-1}$, corresponding to a physical scale of $\sim 1064$ pc—roughly three times the halo's virial radius $R_{\rm vir}$—marking the turbulence driving scale where turbulence energy is injected during the halo assembly. \rev{The turbulence driving scale in our simulations corresponds to $3R_{\rm vir}$, which represents the scale at which turbulent energy is initially injected into the halo. Based on \citet{cuesta08_haloraidus, fong21_haloradius, gao23_haloradius}, $3R_{\rm vir}$ is referred to as the \textit{depletion radius}, marking the outer boundary of the orbiting halo component. The region extending from the halo center to $3R_{\rm vir}$ is of particular interest, as it encompasses complex dynamics where gas flows are strongly affected by the interplay between infalling material and splashback components of both gas and dark matter. These interactions are key drivers of turbulence on halo-wide scales.
} Inside $3R_{\rm vir}$, the inertial range emerges, where turbulence cascades down to smaller scales without energy loss until reaching the dissipation scale, where thermal processes dissipate the energy. The energy spectrum within the inertial scale follows a $k^{-5/3}$ till the dissipation scale. The physical dissipation scale occurs at the atomic level \citep{yuen2022turbulentuniversalgalactickolmogorov}, which is not resolved in our simulations; the numerical viscosity at the particle scale effectively serves as the dissipation mechanism.

\begin{figure}[tbh]
\centering
\includegraphics[width=0.5\columnwidth]{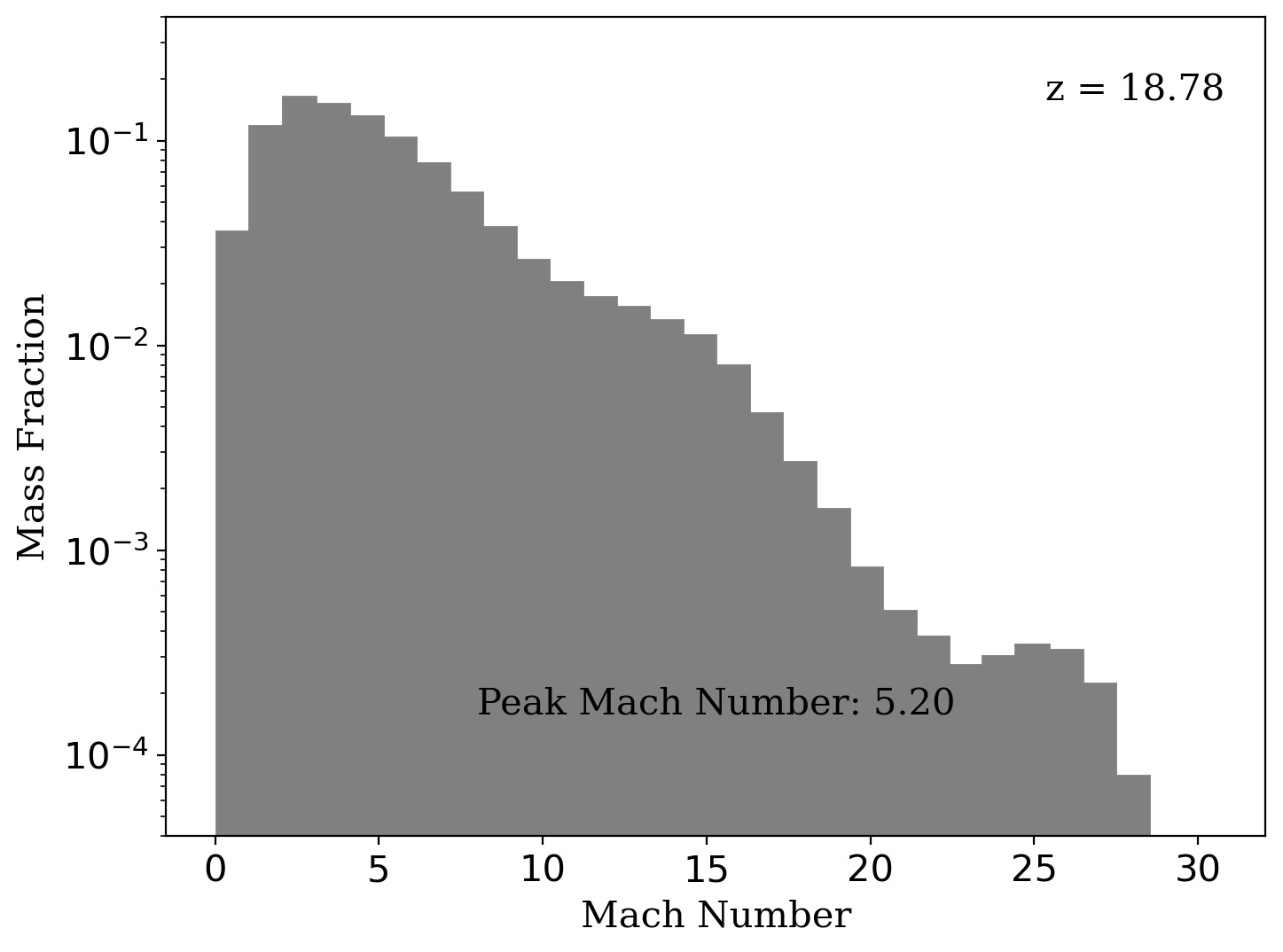}
\caption{The gas mass fraction as a function of \Mach. It shows a peak at \Mach\ $\sim 5.2$, with a broad distribution spanning \Mach\ $\sim 0.5$ to $28$.}
\label{fig:Mach}
\end{figure}

\subsection{Impact of supersonic turbulence on the first star-cloud}

We now examine the characteristic \Mach\ of turbulence in our simulations. Figure~\ref{fig:Mach} shows the fraction of gas mass as a function of \Mach. The distribution peaks at \Mach\ $\approx 5$, with an extended tail reaching up to $\sim 28$. This suggests that most of the halo gas is supersonic, with some fraction of gas of \Mach\ $>10$. Unlike subsonic or transonic flows, supersonic turbulence can compress and shock the gas, leading to the formation of clumpy structures such as giant molecular clouds in the Milky Way. When some of these compressed clumps form dense cores, they can exceed the Jeans mass and collapse to form stars.

\begin{figure}
   \centering
\includegraphics[width=.6\textwidth]{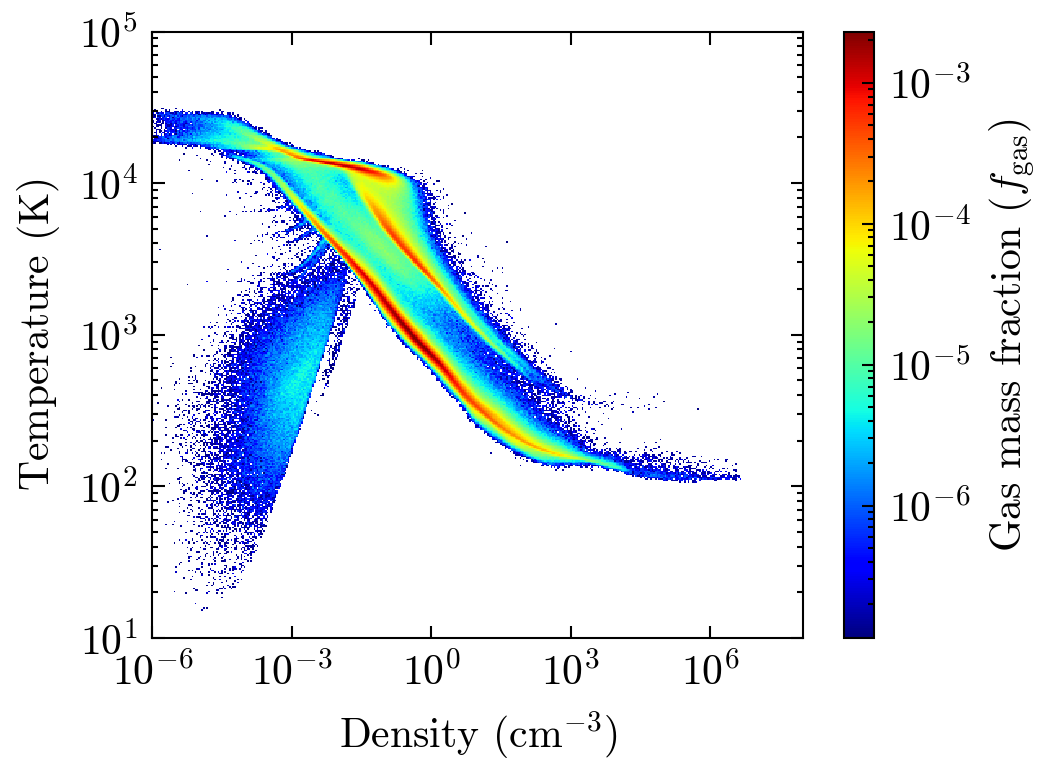}
\caption{ Gas temperature–density phase diagram at the end of the simulations. The gas density spans from \(10^{-5}\) to \(10^{6}\)~\NumDen, corresponding to temperatures ranging from \(10\) to \(10^4\)~K. For \(10^{-5} < n < 10^{-2}\)~\NumDen, the temperature increases with density following adiabatic compression due to gas clustering via gravity. In contrast, for \(n > 10^{-2}\)~\NumDen, the temperature decreases as density increases, owing to efficient cooling by H$_2$ and HD. The warm gas with \(T \sim 10^4\)~K originates from shock-heated diffuse gas in the outer regions of the halo.
}
\label{fig:phase}
\end{figure}

Figure~\ref{fig:phase} shows the phase diagram of the halo gas. Most of the gas resides at densities of $n \approx 10^{-2} - 10^{2}$ \NumDen\ and temperatures between 200 and 2000 K. Only a tiny fraction of the gas exceeds a density of $10^5$ \NumDen, a typical threshold for primordial gas to collapse to form Pop~III stars \citep{op01}. These high-density regions are located at the cores of the densest clumps, where Jeans instability sets in.

\begin{figure}[tbh]
\centering
\includegraphics[width=.6\columnwidth]{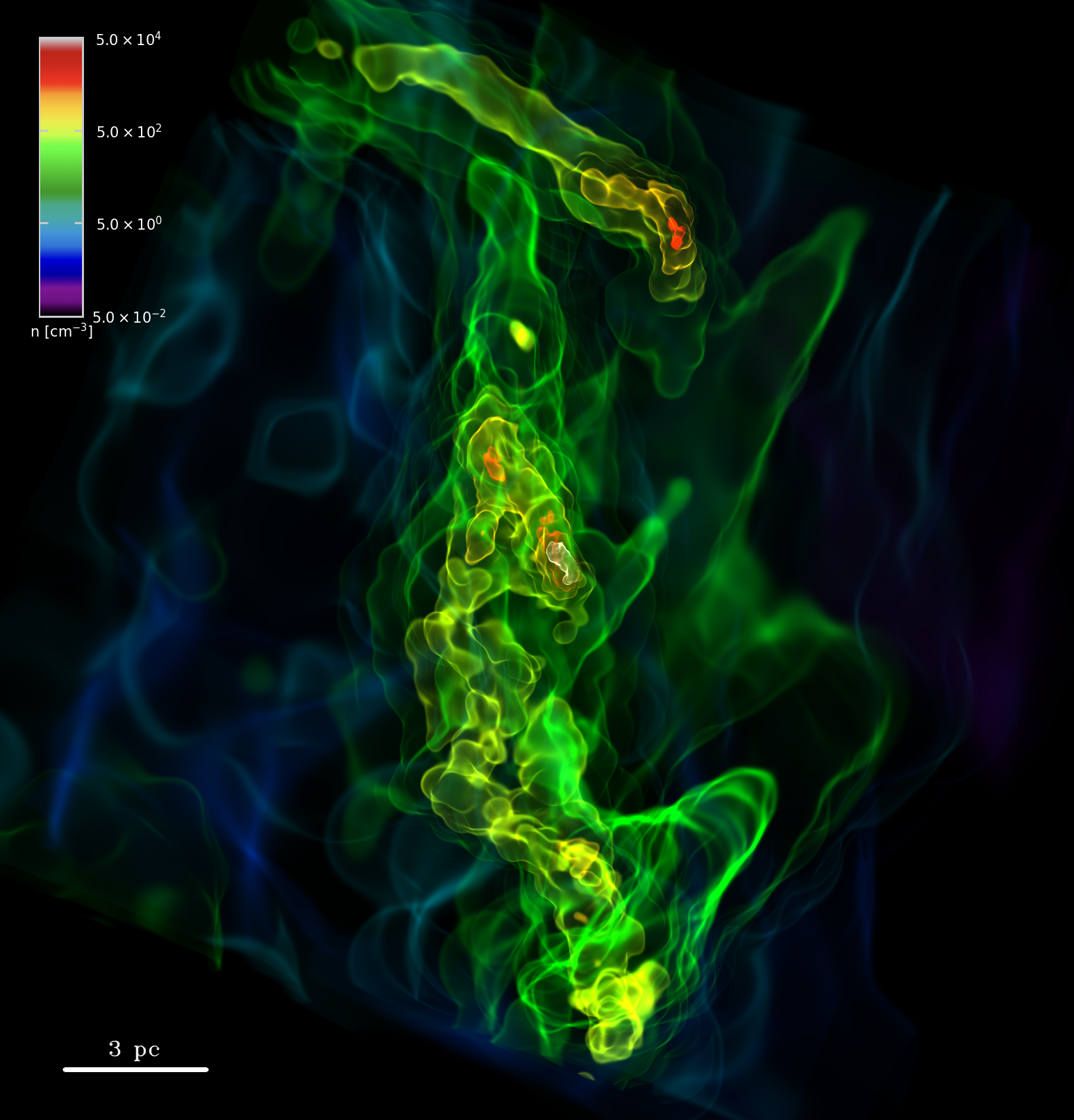}
\caption{3D volume rendering of dense clumps at the center of the halo. Several dense clumps appear as yellow–red nuggets. At this stage, one of the clumps has surpassed the Jeans instability threshold and began to collapse into a Pop~III star with a mass of approximately $8.07\,\Ms$. }
\label{fig:clumps}
\end{figure}

To characterize these clumps, we apply the clump-finding algorithm developed by \citet{Smith2009, 2009Turk} to identify gas clumps within the central minihalo. We found five clumps, but only one is gravitationally bound. This bound core has a mass of $8.07\,\Ms$, a size of 0.03 pc, a density of $3.9 \times 10^6$ \NumDen, and a temperature of 117 K. The core mass already exceeds its Jeans mass of $1.75\,\Ms$ and begins to collapse. Figure~\ref{fig:clumps} shows the structure of the densest clumps, which exhibit irregular and elongated shapes. Other unbound clumps may later become gravitationally bound if the simulation could proceed for a longer time. The clump mass effectively sets an upper limit for the mass of the stars that form. The formation of multiple clumps at the cloud scale implies that the stellar masses will be lower compared to monolithic collapse.

\section{Discussions}

Previous simulations of the first star formation \citep[e.g.,][]{abel1998formationfragmentationprimordialmolecular_PopIII_0, Bromm_2013_The_first_star, Hirano_2014_PopIII_3} typically found that turbulence in primordial star-forming gas was subsonic and had little influence on star formation. However, these studies often focused solely on resolving the innermost regions of minihalos via extremely high levels of hierarchical zoom-in, using AMR or SPH techniques. Consequently, they lacked sufficient resolution at larger scales to properly capture the accretion flows and the development of turbulence. In contrast, our simulations successfully resolve the formation of turbulence during the minihalo assembly. We find that the halo gas is dominated by supersonic turbulence, with a characteristic \Mach\ of $\sim 5-6$. Recently, \citet{tang24} used simulations of driven supersonic turbulence to study its impact on the first star-forming cloud. Their results suggest that supersonic turbulence of \Mach $> 4$ can prevent monolithic collapse in star-forming clouds, instead promoting fragmentation into multiple clumps. Consequently, the typical mass of Pop~III stars would decrease due to turbulence-driven fragmentation. The number of clumps and their relation to the strength of turbulence, as well as the halo's mass, spin, and environment, are promising topics for future exploration. Moreover, supersonic turbulence could amplify magnetic fields via small-scale dynamos, a strong magnetic field might affect the first star formation suggested by \citet{fed11_turb,sharda20, mckee20, stacy22, higashi24_turbmag}.

\citet{Tseliakhovich_2010} suggested that the streaming velocity between gas and dark matter—originating from baryon acoustic oscillations (BAO) in the early universe—can be supersonic. Subsequent studies \citep{hir17, chiou_supersonic, naka22_supersonic} have shown that this streaming velocity can promote the formation of supermassive stars, which may seed early supermassive black holes. However, the effect of streaming velocity is not included in our simulation. Therefore, the supersonic motions observed in our minihalo primarily result from gas accretion and cooling during halo assembly, rather than baryon–dark matter streaming from BAO. Moreover, the supersonic turbulence in our simulations tends to fragment the primordial gas into clumpy structures, favoring the formation of lower-mass Pop.~III stars rather than supermassive stars.

\rev{
Recent high-resolution studies incorporating increasingly accurate gas physics have begun to reveal the fragmentation of protostellar disks during Pop~III star formation \citep{get11, clark11, susa14, riaz18, susa19, wollenberg20, chon21_star, gen22_fragment, higashi22_turb_growth, higashi22_turb, liu24_pop3}. These findings suggest that the characteristic mass of Pop~III stars may shift toward lower values, typically in the range of $0.1$–$100\,\Ms$, and that low-mass stars with $< 1\,\Ms$ may have formed and survived to the present day. For example, \citet{latif22_firststar} performed high-resolution simulations of Pop~III star formation within a small cosmological box of side length $300\, h^{-1}\,\mathrm{ckpc}$, including radiative feedback effects. They found that multiple stars with characteristic masses of $1$–$40\,\Ms$ can form in minihalos, influenced by ionizing UV radiation from massive stars and ejections triggered by three-body interactions. In a separate study, \citet{prole22_firsts} investigated the impact of sink particle implementation on Pop~III star formation by simulating the collapse of a primordial star-forming cloud. Their results indicated that Pop~III stars could have characteristic masses below $1\,\Ms$, and they argued that many earlier studies using sink particles may have overestimated stellar masses and underestimated the number of stars formed.
Although these two studies adopted different modeling approaches, both support the idea that Pop~III stars may typically form with masses below $10\,\Ms$, and that $20$–$70\%$ of them could be dynamically ejected from their host halos shortly after formation. Our results further support this scenario: the supersonic turbulence we find on halo scales leads to fragmentation of the star-forming cloud down to parsec scales. This breaks down the parent gas cloud into smaller clumps, consistent with the formation of lower-mass Pop~III stars as suggested by \citet{latif22_firststar, prole22_firsts}. However, our simulations do not employ sink particles to explicitly follow star formation. Since the final masses of Pop~III stars are sensitive to sink particle treatment, stellar disk evolution, and accretion disk physics on AU scales—processes beyond the scope of this study—we can only infer that Pop~III star masses are bounded by the masses of the collapsing clumps identified in our simulations.
} 
 
Theoretical models predict that Pop~III stars with masses of $80-260\,\Ms$ would die as pair-instability supernovae (PISNe) \citep{brk67,fwh01, hw02,wbh07, joggerst10, hw10, kasen11, chen14c, chen14a, woos17}. If the typical mass scale of Pop~III stars falls below $80\,\Ms$, then stars of $50-80\,\Ms$ may collapse directly into black holes or explode as hypernovae \citep{woo93, mw99, zwh08, Tominaga2009}. Stars of $10-50 \,\Ms$ would die as core-collapse SNe, leaving behind compact remnants such as neutron stars or black holes \citep{woosley_rmp,janka12_ccsne}.

Observations of extremely metal-poor (EMP) stars, believed to bear the chemical signatures of the first SNe, can constrain the initial mass function (IMF) of Pop~III stars \citep{bc05, frebel10, aoki14, ish18, chiaki18, chiaki20, mag20, abe21, 2024Chen}. Reproducing EMP abundance patterns requires strong fallback during SN explosions, achievable either through asymmetry explosions of hypernovae \citep{Umeda&Nomoto2002, maeda03, nom06, Tominaga2007, 2024A&A...681A..44S_EMP_3} or strong fallback of low-energy SNe \citep{keller14, chen17a}. The unique abundance signatures of PISNe, expected from very massive first stars, have not been observed \citep{hw02, hw10, xing23}. 

\rev{Our results highlight the importance of cloud-scale turbulence in minihalos, complementing previous studies that emphasized turbulence in Pop~III protostellar disks. Both cloud-scale fragmentation driven by supersonic turbulence and disk-scale fragmentation induced by disk instability can naturally reduce the characteristic mass of Pop~III stars to the range of $1$–$40\,\Ms$. This provides a compelling explanation for the abundance patterns observed in EMP stars.
}

\section*{Conclusions}

We present the first cosmological zoom-in simulations that successfully resolve turbulent gas accretion during the formation of a massive minihalo of $1.05\times10^7\,\Ms$. Unlike previous studies, our high-resolution \GIZMO\ simulations adopt initial conditions from the state-of-the-art cosmological \TNG\ simulations, enhanced with a robust particle-splitting technique. We identify the emergence of supersonic turbulence within a minihalo, driven by the gravitational infall of gas into the dark matter potential well. The characteristic \Mach\ of the primordial gas reaches approximately 5.2, with a broad distribution spanning from 0.5 to 28. This turbulence fragments the primordial star-forming cloud into multiple clumps. Among them, we find a gravitationally bound core with a mass of $8.07\,\Ms$, exceeding the Jeans threshold, and began to collapse into a Pop~III star. Our results suggest that early structure formation can naturally generate supersonic turbulence, which plays a crucial role in shaping primordial gas clouds and regulating the mass scale of Pop~III stars.
 
\acknowledgments
This research is supported by the National Science and Technology Council of Taiwan, under grant No. MOST 110-2112-M-001-068-MY3, NSTC 113-2112-M-001-028-, 114-2112-M-001-012-, and the Academia Sinica, Taiwan, under a career development award under grant No. AS-CDA-111-M04. KC acknowledges the support of the Alexander von Humboldt Foundation and Heidelberg Institute for Theoretical Studies. This research was supported in part by grant NSF PHY-2309135 to the Kavli Institute for Theoretical Physics (KITP) and grant NSF PHY-2210452 to the Aspen Center for Physics. Our computing resources were supported by the National Energy Research Scientific Computing Center (NERSC), a U.S. Department of Energy Office of Science User Facility operated under Contract No. DE-AC02-05CH11231 and the TIARA Cluster at the Academia Sinica Institute of Astronomy and Astrophysics (ASIAA).



\begin{thebibliography}{}
\providecommand\natexlab[1]{#1}
\providecommand\JournalTitle[1]{#1}

\bibitem[{{Abe} {et~al.}(2021){Abe}, {Yajima}, {Khochfar}, {Dalla Vecchia}, \& {Omukai}}]{abe21}
{Abe}, M., {Yajima}, H., {Khochfar}, S., {Dalla Vecchia}, C., \& {Omukai}, K. 2021, \href{http://dx.doi.org/10.1093/mnras/stab2637}{\JournalTitle{\mnras}, 508, 3226}

\bibitem[{Abel {et~al.}(1998)Abel, Bryan, \& Norman}]{abel1998formationfragmentationprimordialmolecular_PopIII_0}
Abel, T., Bryan, G.~L., \& Norman, M.~L. 1998, The Formation and Fragmentation of Primordial Molecular Clouds, \href{http://arxiv.org/abs/astro-ph/9810215}{{\sffamily arXiv:astro-ph/9810215 [astro-ph]}}

\bibitem[{{Abel} {et~al.}(2002){Abel}, {Bryan}, \& {Norman}}]{abn02}
{Abel}, T., {Bryan}, G.~L., \& {Norman}, M.~L. 2002, \href{http://dx.doi.org/10.1126/science.1063991}{\JournalTitle{Science}, 295, 93}

\bibitem[{{Angl{\'e}s-Alc{\'a}zar} {et~al.}(2021){Angl{\'e}s-Alc{\'a}zar}, {Quataert}, {Hopkins}, {Somerville}, {Hayward}, {Faucher-Gigu{\`e}re}, {Bryan}, {Kere{\v{s}}}, {Hernquist}, \& {Stone}}]{alc21_split}
{Angl{\'e}s-Alc{\'a}zar}, D., {Quataert}, E., {Hopkins}, P.~F., {et~al.} 2021, \href{http://dx.doi.org/10.3847/1538-4357/ac09e8}{\JournalTitle{\apj}, 917, 53}

\bibitem[{{Aoki} {et~al.}(2014){Aoki}, {Tominaga}, {Beers}, {Honda}, \& {Lee}}]{aoki14}
{Aoki}, W., {Tominaga}, N., {Beers}, T.~C., {Honda}, S., \& {Lee}, Y.~S. 2014, \href{http://dx.doi.org/10.1126/science.1252633}{\JournalTitle{Science}, 345, 912}

\bibitem[{{Bacchini} {et~al.}(2020){Bacchini}, {Fraternali}, {Iorio}, {Pezzulli}, {Marasco}, \& {Nipoti}}]{bac20sn}
{Bacchini}, C., {Fraternali}, F., {Iorio}, G., {et~al.} 2020, \href{http://dx.doi.org/10.1051/0004-6361/202038223}{\JournalTitle{\aap}, 641, A70}

\bibitem[{{Barkat} {et~al.}(1967){Barkat}, {Rakavy}, \& {Sack}}]{brk67}
{Barkat}, Z., {Rakavy}, G., \& {Sack}, N. 1967, \href{http://dx.doi.org/10.1103/PhysRevLett.18.379}{\JournalTitle{Physical Review Letters}, 18, 379}

\bibitem[{{Beers} \& {Christlieb}(2005)}]{bc05}
{Beers}, T.~C., \& {Christlieb}, N. 2005, \href{http://dx.doi.org/10.1146/annurev.astro.42.053102.134057}{\JournalTitle{\araa}, 43, 531}

\bibitem[{Bromm(2013)}]{Bromm_2013_The_first_star}
Bromm, V. 2013, \href{http://dx.doi.org/10.1088/0034-4885/76/11/112901}{\JournalTitle{Reports on Progress in Physics}, 76, 112901}

\bibitem[{{Bromm} \& {Larson}(2004)}]{bl04c}
{Bromm}, V., \& {Larson}, R.~B. 2004, \href{http://dx.doi.org/10.1146/annurev.astro.42.053102.134034}{\JournalTitle{\araa}, 42, 79}

\bibitem[{{Chen} {et~al.}(2015){Chen}, {Bromm}, {Heger}, {Jeon}, \& {Woosley}}]{chen15}
{Chen}, K.-J., {Bromm}, V., {Heger}, A., {Jeon}, M., \& {Woosley}, S. 2015, \href{http://dx.doi.org/10.1088/0004-637X/802/1/13}{\JournalTitle{\apj}, 802, 13}

\bibitem[{{Chen} {et~al.}(2017){Chen}, {Heger}, {Whalen}, {Moriya}, {Bromm}, \& {Woosley}}]{chen17a}
{Chen}, K.-J., {Heger}, A., {Whalen}, D.~J., {et~al.} 2017, \href{http://dx.doi.org/10.1093/mnras/stx470}{\JournalTitle{\mnras}, 467, 4731}

\bibitem[{{Chen} {et~al.}(2014{\natexlab{a}}){Chen}, {Heger}, {Woosley}, {Almgren}, \& {Whalen}}]{chen14c}
{Chen}, K.-J., {Heger}, A., {Woosley}, S., {Almgren}, A., \& {Whalen}, D.~J. 2014{\natexlab{a}}, \href{http://dx.doi.org/10.1088/0004-637X/792/1/44}{\JournalTitle{\apj}, 792, 44}

\bibitem[{{Chen} {et~al.}(2024){Chen}, {Tang}, {Whalen}, {Ho}, {Tsai}, {Ou}, \& {Ono}}]{2024Chen}
{Chen}, K.-J., {Tang}, C.-Y., {Whalen}, D.~J., {et~al.} 2024, \href{http://dx.doi.org/10.3847/1538-4357/ad2684}{\JournalTitle{\apj}, 964, 91}

\bibitem[{{Chen} {et~al.}(2014{\natexlab{b}}){Chen}, {Woosley}, {Heger}, {Almgren}, \& {Whalen}}]{chen14a}
{Chen}, K.-J., {Woosley}, S., {Heger}, A., {Almgren}, A., \& {Whalen}, D.~J. 2014{\natexlab{b}}, \href{http://dx.doi.org/10.1088/0004-637X/792/1/28}{\JournalTitle{\apj}, 792, 28}

\bibitem[{{Chiaki} {et~al.}(2018){Chiaki}, {Susa}, \& {Hirano}}]{chiaki18}
{Chiaki}, G., {Susa}, H., \& {Hirano}, S. 2018, \href{http://dx.doi.org/10.1093/mnras/sty040}{\JournalTitle{\mnras}, 475, 4378}

\bibitem[{{Chiaki} \& {Tominaga}(2020)}]{chiaki20}
{Chiaki}, G., \& {Tominaga}, N. 2020, \href{http://dx.doi.org/10.1093/mnras/staa2340}{\JournalTitle{\mnras}, 498, 2676}

\bibitem[{{Chiaki} \& {Yoshida}(2022)}]{gen22_fragment}
{Chiaki}, G., \& {Yoshida}, N. 2022, \href{http://dx.doi.org/10.1093/mnras/stab2799}{\JournalTitle{\mnras}, 510, 5199}

\bibitem[{{Chiou} {et~al.}(2021){Chiou}, {Naoz}, {Burkhart}, {Marinacci}, \& {Vogelsberger}}]{chiou_supersonic}
{Chiou}, Y.~S., {Naoz}, S., {Burkhart}, B., {Marinacci}, F., \& {Vogelsberger}, M. 2021, \href{http://dx.doi.org/10.3847/1538-4357/abc88f}{\JournalTitle{\apj}, 906, 25}

\bibitem[{{Chon} {et~al.}(2021){Chon}, {Omukai}, \& {Schneider}}]{chon21_star}
{Chon}, S., {Omukai}, K., \& {Schneider}, R. 2021, \href{http://dx.doi.org/10.1093/mnras/stab2497}{\JournalTitle{\mnras}, 508, 4175}

\bibitem[{{Clark} {et~al.}(2011){Clark}, {Glover}, {Smith}, {Greif}, {Klessen}, \& {Bromm}}]{clark11}
{Clark}, P.~C., {Glover}, S.~C.~O., {Smith}, R.~J., {et~al.} 2011, \href{http://dx.doi.org/10.1126/science.1198027}{\JournalTitle{Science}, 331, 1040}

\bibitem[{{Cuesta} {et~al.}(2008){Cuesta}, {Prada}, {Klypin}, \& {Moles}}]{cuesta08_haloraidus}
{Cuesta}, A.~J., {Prada}, F., {Klypin}, A., \& {Moles}, M. 2008, \href{http://dx.doi.org/10.1111/j.1365-2966.2008.13590.x}{\JournalTitle{\mnras}, 389, 385}

\bibitem[{{Elmegreen} \& {Scalo}(2004)}]{turb1}
{Elmegreen}, B.~G., \& {Scalo}, J. 2004, \href{http://dx.doi.org/10.1146/annurev.astro.41.011802.094859}{\JournalTitle{\araa}, 42, 211}

\bibitem[{{Falceta-Gon{\c{c}}alves} {et~al.}(2014){Falceta-Gon{\c{c}}alves}, {Kowal}, {Falgarone}, \& {Chian}}]{kowal14}
{Falceta-Gon{\c{c}}alves}, D., {Kowal}, G., {Falgarone}, E., \& {Chian}, A.~C.~L. 2014, \href{http://dx.doi.org/10.5194/npg-21-587-2014}{\JournalTitle{Nonlinear Processes in Geophysics}, 21, 587}

\bibitem[{{Federrath} {et~al.}(2011){Federrath}, {Chabrier}, {Schober}, {Banerjee}, {Klessen}, \& {Schleicher}}]{fed11_turb}
{Federrath}, C., {Chabrier}, G., {Schober}, J., {et~al.} 2011, \href{http://dx.doi.org/10.1103/PhysRevLett.107.114504}{\JournalTitle{\prl}, 107, 114504}

\bibitem[{{Federrath} {et~al.}(2021){Federrath}, {Klessen}, {Iapichino}, \& {Beattie}}]{fed21_turb}
{Federrath}, C., {Klessen}, R.~S., {Iapichino}, L., \& {Beattie}, J.~R. 2021, \href{http://dx.doi.org/10.1038/s41550-020-01282-z}{\JournalTitle{Nature Astronomy}, 5, 365}

\bibitem[{{Fensch} {et~al.}(2023){Fensch}, {Bournaud}, {Brucy}, {Dubois}, {Hennebelle}, \& {Rosdahl}}]{fensch23_gturb}
{Fensch}, J., {Bournaud}, F., {Brucy}, N., {et~al.} 2023, \href{http://dx.doi.org/10.1051/0004-6361/202245491}{\JournalTitle{\aap}, 672, A193}

\bibitem[{{Ferri{\`e}re}(2001)}]{ferri01galaxy}
{Ferri{\`e}re}, K.~M. 2001, \href{http://dx.doi.org/10.1103/RevModPhys.73.1031}{\JournalTitle{Reviews of Modern Physics}, 73, 1031}

\bibitem[{{Fong} \& {Han}(2021)}]{fong21_haloradius}
{Fong}, M., \& {Han}, J. 2021, \href{http://dx.doi.org/10.1093/mnras/stab259}{\JournalTitle{\mnras}, 503, 4250}

\bibitem[{{Frebel}(2010)}]{frebel10}
{Frebel}, A. 2010, \href{http://dx.doi.org/10.1002/asna.201011362}{\JournalTitle{Astronomische Nachrichten}, 331, 474}

\bibitem[{{Fryer} {et~al.}(2001){Fryer}, {Woosley}, \& {Heger}}]{fwh01}
{Fryer}, C.~L., {Woosley}, S.~E., \& {Heger}, A. 2001, \href{http://dx.doi.org/10.1086/319719}{\JournalTitle{\apj}, 550, 372}

\bibitem[{{Gao} {et~al.}(2023){Gao}, {Han}, {Fong}, {Jing}, \& {Li}}]{gao23_haloradius}
{Gao}, H., {Han}, J., {Fong}, M., {Jing}, Y.~P., \& {Li}, Z. 2023, \href{http://dx.doi.org/10.3847/1538-4357/acdfcd}{\JournalTitle{\apj}, 953, 37}

\bibitem[{{Gnedin} {et~al.}(2018){Gnedin}, {Semenov}, \& {Kravtsov}}]{nick18_cfl}
{Gnedin}, N.~Y., {Semenov}, V.~A., \& {Kravtsov}, A.~V. 2018, \href{http://dx.doi.org/10.1016/j.jcp.2018.01.008}{\JournalTitle{Journal of Computational Physics}, 359, 93}

\bibitem[{{Greif}(2015)}]{greif14}
{Greif}, T.~H. 2015, \href{http://dx.doi.org/10.1186/s40668-014-0006-2}{\JournalTitle{Computational Astrophysics and Cosmology}, 2, 3}

\bibitem[{{Greif} {et~al.}(2011){Greif}, {Springel}, {White}, {Glover}, {Clark}, {Smith}, {Klessen}, \& {Bromm}}]{get11}
{Greif}, T.~H., {Springel}, V., {White}, S.~D.~M., {et~al.} 2011, \href{http://dx.doi.org/10.1088/0004-637X/737/2/75}{\JournalTitle{\apj}, 737, 75}

\bibitem[{{Heger} \& {Woosley}(2002)}]{hw02}
{Heger}, A., \& {Woosley}, S.~E. 2002, \href{http://dx.doi.org/10.1086/338487}{\JournalTitle{\apj}, 567, 532}

\bibitem[{{Heger} \& {Woosley}(2010)}]{hw10}
---. 2010, \JournalTitle{\apj}, 724, 341

\bibitem[{{Hennebelle}(2021)}]{henneb21_gturb}
{Hennebelle}, P. 2021, \href{http://dx.doi.org/10.1051/0004-6361/202141650}{\JournalTitle{\aap}, 655, A3}

\bibitem[{{Higashi} {et~al.}(2021){Higashi}, {Susa}, \& {Chiaki}}]{higashi22_turb_growth}
{Higashi}, S., {Susa}, H., \& {Chiaki}, G. 2021, \href{http://dx.doi.org/10.3847/1538-4357/ac01c7}{\JournalTitle{\apj}, 915, 107}

\bibitem[{{Higashi} {et~al.}(2022){Higashi}, {Susa}, \& {Chiaki}}]{higashi22_turb}
---. 2022, \href{http://dx.doi.org/10.3847/1538-4357/ac9b0c}{\JournalTitle{\apj}, 940, 38}

\bibitem[{{Higashi} {et~al.}(2024){Higashi}, {Susa}, {Federrath}, \& {Chiaki}}]{higashi24_turbmag}
{Higashi}, S., {Susa}, H., {Federrath}, C., \& {Chiaki}, G. 2024, \href{http://dx.doi.org/10.3847/1538-4357/ad2066}{\JournalTitle{\apj}, 962, 158}

\bibitem[{{Hirano} {et~al.}(2017){Hirano}, {Hosokawa}, {Yoshida}, \& {Kuiper}}]{hir17}
{Hirano}, S., {Hosokawa}, T., {Yoshida}, N., \& {Kuiper}, R. 2017, \href{http://dx.doi.org/10.1126/science.aai9119}{\JournalTitle{Science}, 357, 1375}

\bibitem[{{Hirano} {et~al.}(2015){Hirano}, {Hosokawa}, {Yoshida}, {Omukai}, \& {Yorke}}]{hir15}
{Hirano}, S., {Hosokawa}, T., {Yoshida}, N., {Omukai}, K., \& {Yorke}, H.~W. 2015, \href{http://dx.doi.org/10.1093/mnras/stv044}{\JournalTitle{\mnras}, 448, 568}

\bibitem[{{Hirano} {et~al.}(2014){Hirano}, {Hosokawa}, {Yoshida}, {Umeda}, {Omukai}, {Chiaki}, \& {Yorke}}]{hir13}
{Hirano}, S., {Hosokawa}, T., {Yoshida}, N., {et~al.} 2014, \href{http://dx.doi.org/10.1088/0004-637X/781/2/60}{\JournalTitle{\apj}, 781, 60}

\bibitem[{Hirano {et~al.}(2014)Hirano, Hosokawa, Yoshida, Umeda, Omukai, Chiaki, \& Yorke}]{Hirano_2014_PopIII_3}
Hirano, S., Hosokawa, T., Yoshida, N., {et~al.} 2014, \href{http://dx.doi.org/10.1088/0004-637x/781/2/60}{\JournalTitle{The Astrophysical Journal}, 781, 60}

\bibitem[{Hopkins(2015)}]{Hopkins_2015}
Hopkins, P.~F. 2015, \href{http://dx.doi.org/10.1093/mnras/stv195}{\JournalTitle{Monthly Notices of the Royal Astronomical Society}, 450, 53–110}

\bibitem[{Hopkins {et~al.}(2018)Hopkins, Wetzel, Kereš, Faucher-Giguère, Quataert, Boylan-Kolchin, Murray, Hayward, Garrison-Kimmel, Hummels, Feldmann, Torrey, Ma, Anglés-Alcázar, Su, Orr, Schmitz, Escala, Sanderson, Grudić, Hafen, Kim, Fitts, Bullock, Wheeler, Chan, Elbert, \& Narayanan}]{Hopkins_2018_cooling}
Hopkins, P.~F., Wetzel, A., Kereš, D., {et~al.} 2018, \href{http://dx.doi.org/10.1093/mnras/sty1690}{\JournalTitle{Monthly Notices of the Royal Astronomical Society}, 480, 800–863}

\bibitem[{{Hu} {et~al.}(2022){Hu}, {Qiu}, {Gendron-Marsolais}, {Bogdanovi{\'c}}, {Hlavacek-Larrondo}, {Ho}, {Inayoshi}, \& {McNamara}}]{hu22turb}
{Hu}, H., {Qiu}, Y., {Gendron-Marsolais}, M.-L., {et~al.} 2022, \href{http://dx.doi.org/10.3847/2041-8213/ac6601}{\JournalTitle{\apjl}, 929, L30}

\bibitem[{{Ishigaki} {et~al.}(2018){Ishigaki}, {Tominaga}, {Kobayashi}, \& {Nomoto}}]{ish18}
{Ishigaki}, M.~N., {Tominaga}, N., {Kobayashi}, C., \& {Nomoto}, K. 2018, \href{http://dx.doi.org/10.3847/1538-4357/aab3de}{\JournalTitle{\apj}, 857, 46}

\bibitem[{{Janka}(2012)}]{janka12_ccsne}
{Janka}, H.-T. 2012, \href{http://dx.doi.org/10.1146/annurev-nucl-102711-094901}{\JournalTitle{Annual Review of Nuclear and Particle Science}, 62, 407}

\bibitem[{{Joggerst} {et~al.}(2010){Joggerst}, {Almgren}, {Bell}, {Heger}, {Whalen}, \& {Woosley}}]{joggerst10}
{Joggerst}, C.~C., {Almgren}, A., {Bell}, J., {et~al.} 2010, \href{http://dx.doi.org/10.1088/0004-637X/709/1/11}{\JournalTitle{\apj}, 709, 11}

\bibitem[{{Kasen} {et~al.}(2011){Kasen}, {Woosley}, \& {Heger}}]{kasen11}
{Kasen}, D., {Woosley}, S.~E., \& {Heger}, A. 2011, \href{http://dx.doi.org/10.1088/0004-637X/734/2/102}{\JournalTitle{\apj}, 734, 102}

\bibitem[{{Keller} {et~al.}(2014){Keller}, {Bessell}, {Frebel}, {Casey}, {Asplund}, {Jacobson}, {Lind}, {Norris}, {Yong}, {Heger}, {Magic}, {da Costa}, {Schmidt}, \& {Tisserand}}]{keller14}
{Keller}, S.~C., {Bessell}, M.~S., {Frebel}, A., {et~al.} 2014, \href{http://dx.doi.org/10.1038/nature12990}{\JournalTitle{\nat}, 506, 463}

\bibitem[{{Kim} \& {Ostriker}(2017)}]{kim17sn}
{Kim}, C.-G., \& {Ostriker}, E.~C. 2017, \href{http://dx.doi.org/10.3847/1538-4357/aa8599}{\JournalTitle{\apj}, 846, 133}

\bibitem[{{Klessen} \& {Glover}(2023)}]{ks23}
{Klessen}, R.~S., \& {Glover}, S. C.~O. 2023, \href{http://dx.doi.org/10.1146/annurev-astro-071221-053453}{\JournalTitle{\araa}, 61, 65}

\bibitem[{{Klessen} {et~al.}(2000){Klessen}, {Heitsch}, \& {Mac Low}}]{klessen00turb}
{Klessen}, R.~S., {Heitsch}, F., \& {Mac Low}, M.-M. 2000, \href{http://dx.doi.org/10.1086/308891}{\JournalTitle{\apj}, 535, 887}

\bibitem[{Kolmogorov(1991)}]{1991_Kolmogorov}
Kolmogorov, A.~N. 1991, \href{http://www.jstor.org/stable/51980}{\JournalTitle{Proceedings: Mathematical and Physical Sciences}, 434, 9}

\bibitem[{{Koo} {et~al.}(2020){Koo}, {Kim}, {Park}, \& {Ostriker}}]{koo20sn}
{Koo}, B.-C., {Kim}, C.-G., {Park}, S., \& {Ostriker}, E.~C. 2020, \href{http://dx.doi.org/10.3847/1538-4357/abc1e7}{\JournalTitle{\apj}, 905, 35}

\bibitem[{{Krumholz} \& {Burkhart}(2016)}]{kru16feedback}
{Krumholz}, M.~R., \& {Burkhart}, B. 2016, \href{http://dx.doi.org/10.1093/mnras/stw434}{\JournalTitle{\mnras}, 458, 1671}

\bibitem[{{Larson}(1979)}]{larson79turb}
{Larson}, R.~B. 1979, \href{http://dx.doi.org/10.1093/mnras/186.3.479}{\JournalTitle{\mnras}, 186, 479}

\bibitem[{{Latif} {et~al.}(2022){Latif}, {Whalen}, \& {Khochfar}}]{latif22_firststar}
{Latif}, M.~A., {Whalen}, D., \& {Khochfar}, S. 2022, \href{http://dx.doi.org/10.3847/1538-4357/ac3916}{\JournalTitle{\apj}, 925, 28}

\bibitem[{{Liu} {et~al.}(2024){Liu}, {Gurian}, {Inayoshi}, {Hirano}, {Hosokawa}, {Bromm}, \& {Yoshida}}]{liu24_pop3}
{Liu}, B., {Gurian}, J., {Inayoshi}, K., {et~al.} 2024, \href{http://dx.doi.org/10.1093/mnras/stae2066}{\JournalTitle{\mnras}, 534, 290}

\bibitem[{{Mac Low} \& {Klessen}(2004)}]{low04rmp}
{Mac Low}, M.-M., \& {Klessen}, R.~S. 2004, \href{http://dx.doi.org/10.1103/RevModPhys.76.125}{\JournalTitle{Reviews of Modern Physics}, 76, 125}

\bibitem[{{MacFadyen} \& {Woosley}(1999)}]{mw99}
{MacFadyen}, A.~I., \& {Woosley}, S.~E. 1999, \href{http://dx.doi.org/10.1086/307790}{\JournalTitle{\apj}, 524, 262}

\bibitem[{{Maeda} \& {Nomoto}(2003)}]{maeda03}
{Maeda}, K., \& {Nomoto}, K. 2003, \href{http://dx.doi.org/10.1086/378948}{\JournalTitle{\apj}, 598, 1163}

\bibitem[{{Magg} {et~al.}(2020){Magg}, {Nordlander}, {Glover}, {Hansen}, {Ishigaki}, {Heger}, {Klessen}, {Kobayashi}, \& {Nomoto}}]{mag20}
{Magg}, M., {Nordlander}, T., {Glover}, S. C.~O., {et~al.} 2020, \href{http://dx.doi.org/10.1093/mnras/staa2624}{\JournalTitle{\mnras}, 498, 3703}

\bibitem[{{Marinacci} {et~al.}(2018){Marinacci}, {Vogelsberger}, {Pakmor}, {Torrey}, {Springel}, {Hernquist}, {Nelson}, {Weinberger}, {Pillepich}, {Naiman}, \& {Genel}}]{2018Marinacci}
{Marinacci}, F., {Vogelsberger}, M., {Pakmor}, R., {et~al.} 2018, \href{http://dx.doi.org/10.1093/mnras/sty2206}{\JournalTitle{\mnras}, 480, 5113}

\bibitem[{{McKee} {et~al.}(2020){McKee}, {Stacy}, \& {Li}}]{mckee20}
{McKee}, C.~F., {Stacy}, A., \& {Li}, P.~S. 2020, \href{http://dx.doi.org/10.1093/mnras/staa1903}{\JournalTitle{\mnras}, 496, 5528}

\bibitem[{{Naiman} {et~al.}(2018){Naiman}, {Pillepich}, {Springel}, {Ramirez-Ruiz}, {Torrey}, {Vogelsberger}, {Pakmor}, {Nelson}, {Marinacci}, {Hernquist}, {Weinberger}, \& {Genel}}]{2018Naiman}
{Naiman}, J.~P., {Pillepich}, A., {Springel}, V., {et~al.} 2018, \href{http://dx.doi.org/10.1093/mnras/sty618}{\JournalTitle{\mnras}, 477, 1206}

\bibitem[{{Nakazato} {et~al.}(2022){Nakazato}, {Chiaki}, {Yoshida}, {Naoz}, {Lake}, \& {Chiou}}]{naka22_supersonic}
{Nakazato}, Y., {Chiaki}, G., {Yoshida}, N., {et~al.} 2022, \href{http://dx.doi.org/10.3847/2041-8213/ac573e}{\JournalTitle{\apjl}, 927, L12}

\bibitem[{{Nelson} {et~al.}(2018){Nelson}, {Pillepich}, {Springel}, {Weinberger}, {Hernquist}, {Pakmor}, {Genel}, {Torrey}, {Vogelsberger}, {Kauffmann}, {Marinacci}, \& {Naiman}}]{2018Nelson}
{Nelson}, D., {Pillepich}, A., {Springel}, V., {et~al.} 2018, \href{http://dx.doi.org/10.1093/mnras/stx3040}{\JournalTitle{\mnras}, 475, 624}

\bibitem[{{Nelson} {et~al.}(2019){Nelson}, {Pillepich}, {Springel}, {Pakmor}, {Weinberger}, {Genel}, {Torrey}, {Vogelsberger}, {Marinacci}, \& {Hernquist}}]{2019Nelson}
---. 2019, \href{http://dx.doi.org/10.1093/mnras/stz2306}{\JournalTitle{\mnras}, 490, 3234}

\bibitem[{{Nomoto} {et~al.}(2006){Nomoto}, {Tominaga}, {Umeda}, {Kobayashi}, \& {Maeda}}]{nom06}
{Nomoto}, K., {Tominaga}, N., {Umeda}, H., {Kobayashi}, C., \& {Maeda}, K. 2006, \href{http://dx.doi.org/10.1016/j.nuclphysa.2006.05.008}{\JournalTitle{Nuclear Physics A}, 777, 424}

\bibitem[{{Omukai} \& {Palla}(2001)}]{op01}
{Omukai}, K., \& {Palla}, F. 2001, \href{http://dx.doi.org/10.1086/324410}{\JournalTitle{\apjl}, 561, L55}

\bibitem[{{Parrish} {et~al.}(2010){Parrish}, {Quataert}, \& {Sharma}}]{par10cluster}
{Parrish}, I.~J., {Quataert}, E., \& {Sharma}, P. 2010, \href{http://dx.doi.org/10.1088/2041-8205/712/2/L194}{\JournalTitle{\apjl}, 712, L194}

\bibitem[{{Pillepich} {et~al.}(2018){Pillepich}, {Nelson}, {Hernquist}, {Springel}, {Pakmor}, {Torrey}, {Weinberger}, {Genel}, {Naiman}, {Marinacci}, \& {Vogelsberger}}]{2018Pillepich}
{Pillepich}, A., {Nelson}, D., {Hernquist}, L., {et~al.} 2018, \href{http://dx.doi.org/10.1093/mnras/stx3112}{\JournalTitle{\mnras}, 475, 648}

\bibitem[{{Pillepich} {et~al.}(2019){Pillepich}, {Nelson}, {Springel}, {Pakmor}, {Torrey}, {Weinberger}, {Vogelsberger}, {Marinacci}, {Genel}, {van der Wel}, \& {Hernquist}}]{2019Pillepich}
{Pillepich}, A., {Nelson}, D., {Springel}, V., {et~al.} 2019, \href{http://dx.doi.org/10.1093/mnras/stz2338}{\JournalTitle{\mnras}, 490, 3196}

\bibitem[{{Prole} {et~al.}(2022){Prole}, {Clark}, {Klessen}, \& {Glover}}]{prole22_firsts}
{Prole}, L.~R., {Clark}, P.~C., {Klessen}, R.~S., \& {Glover}, S. C.~O. 2022, \href{http://dx.doi.org/10.1093/mnras/stab3697}{\JournalTitle{\mnras}, 510, 4019}

\bibitem[{{Ramesh} \& {Nelson}(2024)}]{2024Ramesh_1}
{Ramesh}, R., \& {Nelson}, D. 2024, \href{http://dx.doi.org/10.1093/mnras/stae237}{\JournalTitle{\mnras}, 528, 3320}

\bibitem[{{Ramesh} {et~al.}(2024){Ramesh}, {Nelson}, {Fielding}, \& {Br{\"u}ggen}}]{2024Ramesh_2}
{Ramesh}, R., {Nelson}, D., {Fielding}, D., \& {Br{\"u}ggen}, M. 2024, \href{http://dx.doi.org/10.1051/0004-6361/202348786}{\JournalTitle{\aap}, 684, L16}

\bibitem[{{Riaz} {et~al.}(2018){Riaz}, {Bovino}, {Vanaverbeke}, \& {Schleicher}}]{riaz18}
{Riaz}, R., {Bovino}, S., {Vanaverbeke}, S., \& {Schleicher}, D.~R.~G. 2018, \href{http://dx.doi.org/10.1093/mnras/sty1635}{\JournalTitle{\mnras}, 479, 667}

\bibitem[{{Ruszkowski} \& {Oh}(2011)}]{ro11cluster}
{Ruszkowski}, M., \& {Oh}, S.~P. 2011, \href{http://dx.doi.org/10.1111/j.1365-2966.2011.18482.x}{\JournalTitle{\mnras}, 414, 1493}

\bibitem[{{Sharda} {et~al.}(2020){Sharda}, {Federrath}, \& {Krumholz}}]{sharda20}
{Sharda}, P., {Federrath}, C., \& {Krumholz}, M.~R. 2020, \href{http://dx.doi.org/10.1093/mnras/staa1926}{\JournalTitle{\mnras}, 497, 336}

\bibitem[{{Shore}(1992)}]{shore92}
{Shore}, S.~N. 1992, {An introduction to astrophysical hydrodynamics}

\bibitem[{{Sk{\'u}lad{\'o}ttir} {et~al.}(2024){Sk{\'u}lad{\'o}ttir}, {Vanni}, {Salvadori}, \& {Lucchesi}}]{2024A&A...681A..44S_EMP_3}
{Sk{\'u}lad{\'o}ttir}, {\'A}., {Vanni}, I., {Salvadori}, S., \& {Lucchesi}, R. 2024, \href{http://dx.doi.org/10.1051/0004-6361/202346231}{\JournalTitle{\aap}, 681, A44}

\bibitem[{Smith {et~al.}(2009)Smith, Turk, Sigurdsson, O'Shea, \& Norman}]{Smith2009}
Smith, B.~D., Turk, M.~J., Sigurdsson, S., O'Shea, B.~W., \& Norman, M.~L. 2009, \href{http://dx.doi.org/10.1088/0004-637X/691/1/441}{\JournalTitle{The Astrophysical Journal}, 691, 441}

\bibitem[{{Smith} {et~al.}(2017){Smith}, {Bryan}, {Glover}, {Goldbaum}, {Turk}, {Regan}, {Wise}, {Schive}, {Abel}, {Emerick}, {O'Shea}, {Anninos}, {Hummels}, \& {Khochfar}}]{2017MNRAS.466.2217S_Grackle}
{Smith}, B.~D., {Bryan}, G.~L., {Glover}, S. C.~O., {et~al.} 2017, \href{http://dx.doi.org/10.1093/mnras/stw3291}{\JournalTitle{\mnras}, 466, 2217}

\bibitem[{{Springel} \& {Hernquist}(2003)}]{Springel2003}
{Springel}, V., \& {Hernquist}, L. 2003, \href{http://dx.doi.org/10.1046/j.1365-8711.2003.06206.x}{\JournalTitle{\mnras}, 339, 289}

\bibitem[{{Springel} {et~al.}(2001){Springel}, {White}, {Tormen}, \& {Kauffmann}}]{2001Springel}
{Springel}, V., {White}, S. D.~M., {Tormen}, G., \& {Kauffmann}, G. 2001, \href{http://dx.doi.org/10.1046/j.1365-8711.2001.04912.x}{\JournalTitle{\mnras}, 328, 726}

\bibitem[{{Springel} {et~al.}(2018){Springel}, {Pakmor}, {Pillepich}, {Weinberger}, {Nelson}, {Hernquist}, {Vogelsberger}, {Genel}, {Torrey}, {Marinacci}, \& {Naiman}}]{2018Springel}
{Springel}, V., {Pakmor}, R., {Pillepich}, A., {et~al.} 2018, \href{http://dx.doi.org/10.1093/mnras/stx3304}{\JournalTitle{\mnras}, 475, 676}

\bibitem[{{Stacy} {et~al.}(2010){Stacy}, {Greif}, \& {Bromm}}]{stacy10}
{Stacy}, A., {Greif}, T.~H., \& {Bromm}, V. 2010, \href{http://dx.doi.org/10.1111/j.1365-2966.2009.16113.x}{\JournalTitle{\mnras}, 403, 45}

\bibitem[{{Stacy} {et~al.}(2022){Stacy}, {McKee}, {Lee}, {Klein}, \& {Li}}]{stacy22}
{Stacy}, A., {McKee}, C.~F., {Lee}, A.~T., {Klein}, R.~I., \& {Li}, P.~S. 2022, \href{http://dx.doi.org/10.1093/mnras/stac372}{\JournalTitle{\mnras}, 511, 5042}

\bibitem[{{Susa}(2019)}]{susa19}
{Susa}, H. 2019, \href{http://dx.doi.org/10.3847/1538-4357/ab1b6f}{\JournalTitle{\apj}, 877, 99}

\bibitem[{{Susa} {et~al.}(2014){Susa}, {Hasegawa}, \& {Tominaga}}]{susa14}
{Susa}, H., {Hasegawa}, K., \& {Tominaga}, N. 2014, \href{http://dx.doi.org/10.1088/0004-637X/792/1/32}{\JournalTitle{\apj}, 792, 32}

\bibitem[{{Tang} \& {Chen}(2024)}]{tang24}
{Tang}, C.-Y., \& {Chen}, K.-J. 2024, \href{http://dx.doi.org/10.1093/mnras/stae764}{\JournalTitle{\mnras}, 529, 4248}

\bibitem[{Tegmark {et~al.}(1997)Tegmark, Silk, Rees, Blanchard, Abel, \& Palla}]{Tegmark_1997}
Tegmark, M., Silk, J., Rees, M.~J., {et~al.} 1997, \href{http://dx.doi.org/10.1086/303434}{\JournalTitle{The Astrophysical Journal}, 474, 1–12}

\bibitem[{{Tominaga}(2009)}]{Tominaga2009}
{Tominaga}, N. 2009, \href{http://dx.doi.org/10.1088/0004-637X/690/1/526}{\JournalTitle{\apj}, 690, 526}

\bibitem[{{Tominaga} {et~al.}(2007){Tominaga}, {Umeda}, \& {Nomoto}}]{Tominaga2007}
{Tominaga}, N., {Umeda}, H., \& {Nomoto}, K. 2007, \href{http://dx.doi.org/10.1086/513063}{\JournalTitle{\apj}, 660, 516}

\bibitem[{Tseliakhovich \& Hirata(2010)}]{Tseliakhovich_2010}
Tseliakhovich, D., \& Hirata, C. 2010, \href{http://dx.doi.org/10.1103/PhysRevD.82.083520}{\JournalTitle{Phys. Rev. D}, 82, 083520}

\bibitem[{{Tung} \& {Chen}(2024)}]{2024Tung}
{Tung}, P.-C., \& {Chen}, K.-J. 2024, \href{http://dx.doi.org/10.48550/arXiv.2412.16440}{\JournalTitle{arXiv e-prints}, arXiv:2412.16440}

\bibitem[{Turk {et~al.}(2009)Turk, Abel, \& O'Shea}]{2009Turk}
Turk, M.~J., Abel, T., \& O'Shea, B. 2009, \JournalTitle{Science}, 325, 601

\bibitem[{{Turk} {et~al.}(2009){Turk}, {Abel}, \& {O'Shea}}]{turk09}
{Turk}, M.~J., {Abel}, T., \& {O'Shea}, B. 2009, \href{http://dx.doi.org/10.1126/science.1173540}{\JournalTitle{Science}, 325, 601}

\bibitem[{{Umeda} \& {Nomoto}(2002)}]{Umeda&Nomoto2002}
{Umeda}, H., \& {Nomoto}, K. 2002, \href{http://dx.doi.org/10.1086/323946}{\JournalTitle{\apj}, 565, 385}

\bibitem[{{Wollenberg} {et~al.}(2020){Wollenberg}, {Glover}, {Clark}, \& {Klessen}}]{wollenberg20}
{Wollenberg}, K. M.~J., {Glover}, S. C.~O., {Clark}, P.~C., \& {Klessen}, R.~S. 2020, \href{http://dx.doi.org/10.1093/mnras/staa289}{\JournalTitle{\mnras}, 494, 1871}

\bibitem[{{Woosley}(1993)}]{woo93}
{Woosley}, S.~E. 1993, \href{http://dx.doi.org/10.1086/172359}{\JournalTitle{\apj}, 405, 273}

\bibitem[{{Woosley}(2017)}]{woos17}
---. 2017, \href{http://dx.doi.org/10.3847/1538-4357/836/2/244}{\JournalTitle{\apj}, 836, 244}

\bibitem[{{Woosley} {et~al.}(2007){Woosley}, {Blinnikov}, \& {Heger}}]{wbh07}
{Woosley}, S.~E., {Blinnikov}, S., \& {Heger}, A. 2007, \href{http://dx.doi.org/10.1038/nature06333}{\JournalTitle{\nat}, 450, 390}

\bibitem[{{Woosley} {et~al.}(2002){Woosley}, {Heger}, \& {Weaver}}]{woosley_rmp}
{Woosley}, S.~E., {Heger}, A., \& {Weaver}, T.~A. 2002, \href{http://dx.doi.org/10.1103/RevModPhys.74.1015}{\JournalTitle{Reviews of Modern Physics}, 74, 1015}

\bibitem[{{Xing} {et~al.}(2023){Xing}, {Zhao}, {Liu}, {Heger}, {Han}, {Aoki}, {Chen}, {Ishigaki}, {Li}, \& {Zhao}}]{xing23}
{Xing}, Q.-F., {Zhao}, G., {Liu}, Z.-W., {et~al.} 2023, \href{http://dx.doi.org/10.1038/s41586-023-06028-1}{\JournalTitle{\nat}, 618, 712}

\bibitem[{{Yoshida} {et~al.}(2008){Yoshida}, {Omukai}, \& {Hernquist}}]{y08}
{Yoshida}, N., {Omukai}, K., \& {Hernquist}, L. 2008, \href{http://dx.doi.org/10.1126/science.1160259}{\JournalTitle{Science}, 321, 669}

\bibitem[{Yuen {et~al.}(2022)Yuen, Ho, Law, Chen, \& Lazarian}]{yuen2022turbulentuniversalgalactickolmogorov}
Yuen, K.~H., Ho, K.~W., Law, C.~Y., Chen, A., \& Lazarian, A. 2022, Turbulent universal galactic Kolmogorov velocity cascade over 6 decades, \href{http://arxiv.org/abs/2204.13760}{{\sffamily arXiv:2204.13760 [astro-ph.GA]}}

\bibitem[{{Zhang} {et~al.}(2008){Zhang}, {Woosley}, \& {Heger}}]{zwh08}
{Zhang}, W., {Woosley}, S.~E., \& {Heger}, A. 2008, \href{http://dx.doi.org/10.1086/526404}{\JournalTitle{\apj}, 679, 639}

\end{thebibliography}

\end{document}